\documentclass[reprint,superscriptaddress]{revtex4-1}

\pdfoutput=1

\usepackage{graphicx}
\usepackage{dcolumn}
\usepackage{bm}
\usepackage{braket}
\usepackage{float}
\usepackage{amsmath}

\begin{document}

\newcommand{\smb}{SmB$_{6}$}
\newcommand{\kt}{k_B T}

\title{Investigation of high-temperature bulk transport characteristics and skew scattering in samarium hexaboride}

\author{Alexa Rakoski}
\email[]{ralexa@umich.edu}
\affiliation{Department of Physics, University of Michigan, Ann Arbor, MI 48109, USA}

\author{Yun Suk Eo}
\affiliation{Department of Physics, University of Michigan, Ann Arbor, MI 48109, USA}
\affiliation{Department of Physics, University of Maryland, College Park, MD 20740, USA}

\author{\c{C}a\u{g}l{\i}yan Kurdak}
\affiliation{Department of Physics, University of Michigan, Ann Arbor, MI 48109, USA}

\author{Boyoun Kang}
\affiliation{Department of Materials Science and Engineering, Gwangju Institute of Science and Technology (GIST), Gwangju 61005 Korea}

\author{Myungsuk Song}
\affiliation{Department of Materials Science and Engineering, Gwangju Institute of Science and Technology (GIST), Gwangju 61005 Korea}

\author{Beongki Cho}
\affiliation{Department of Materials Science and Engineering, Gwangju Institute of Science and Technology (GIST), Gwangju 61005 Korea}

\date{\today}

\begin{abstract}
A well-known feature in transport data of the topological Kondo insulator \smb{} is the sign change in the Hall coefficient at 65 K. Carriers in \smb{} are known to be negative, but above 65 K, the Hall sign suggests that the carriers are positive. Here, we extend Hall measurements up to 400 K and observe that the Hall coefficient changes back to the correct (negative) sign at about 305 K. We interpret the anomalous sign of the Hall coefficient in the context of skew scattering arising from the strong correlations between the $f$ and $d$ electrons. At energy scales where the gap is closed, the number of $d$ electrons in resonance with the $f$ electrons at the Fermi energy varies. When a large proportion of $d$ and $f$ electrons are in resonance, skew scattering is dominant, leading to the observation of the positive sign, but when fewer are in resonance, conventional scattering mechanisms dominate instead. 
\end{abstract}

\maketitle
\nopagebreak

Samarium hexaboride (\smb) is a long-studied compound with a unique combination of characteristics, with one of the earliest studies published 50 years ago in 1969. \cite{menth} Past research led to the classification of \smb{} as a rare-earth mixed valence compound \cite{mott74, martin79, allen80} as well as a Kondo insulator, \cite{aeppli, riseborough00} where strong interactions between $f$ and $d$ electrons lead to the opening of a small hybridization gap at the Fermi energy below about 100 K. In transport, the gap opening is observed as a resistance rise below about 50 K, terminating in a conductive plateau at about 4 K. \cite{menth, allen} Historical attempts made to explain this behavior most commonly attributed it to impurity conduction, \cite{menth, nickerson} but recent theoretical developments have classified \smb{} as the first correlated topological insulator. \cite{dzero10,takimoto,dzero12} Transport experiments have demonstrated that the plateau indeed arises from surface conduction, \cite{wolgast, kim13, kim14} and much of the work on \smb{} in recent years has continued to investigate its surface characteristics. \cite{wolgastmt, zhang, neupane, jiang, nxu14, frantzeskakis, denlingerlinked, liscience,xiang17, rossler16, pirie}

However, the bulk of \smb{} is also interesting in light of recent transport results that bypass the surface conduction and show that the bulk of \smb{} is truly insulating and immune to disorder. \cite{eopnas} This property is quite different from the standard topological insulators like Bi$_2$Se$_3$, where the addition of impurities creates an in-gap bulk state. In \smb, in-gap bulk states are not supported, \cite{rakoski} and the persistence of a clean gap in the presence of impurities can be thought of in analogy to a fully gapped BCS superconductor. \cite{bcs, reif}  In addition, quantum oscillation results suggesting an unconventional bulk Fermi surface in \smb{} indicate that the bulk properties require further investigation. \cite{tan, hartstein}

\begin{figure}
\includegraphics[scale=1, trim = 0mm 0mm 0mm 0mm, clip]{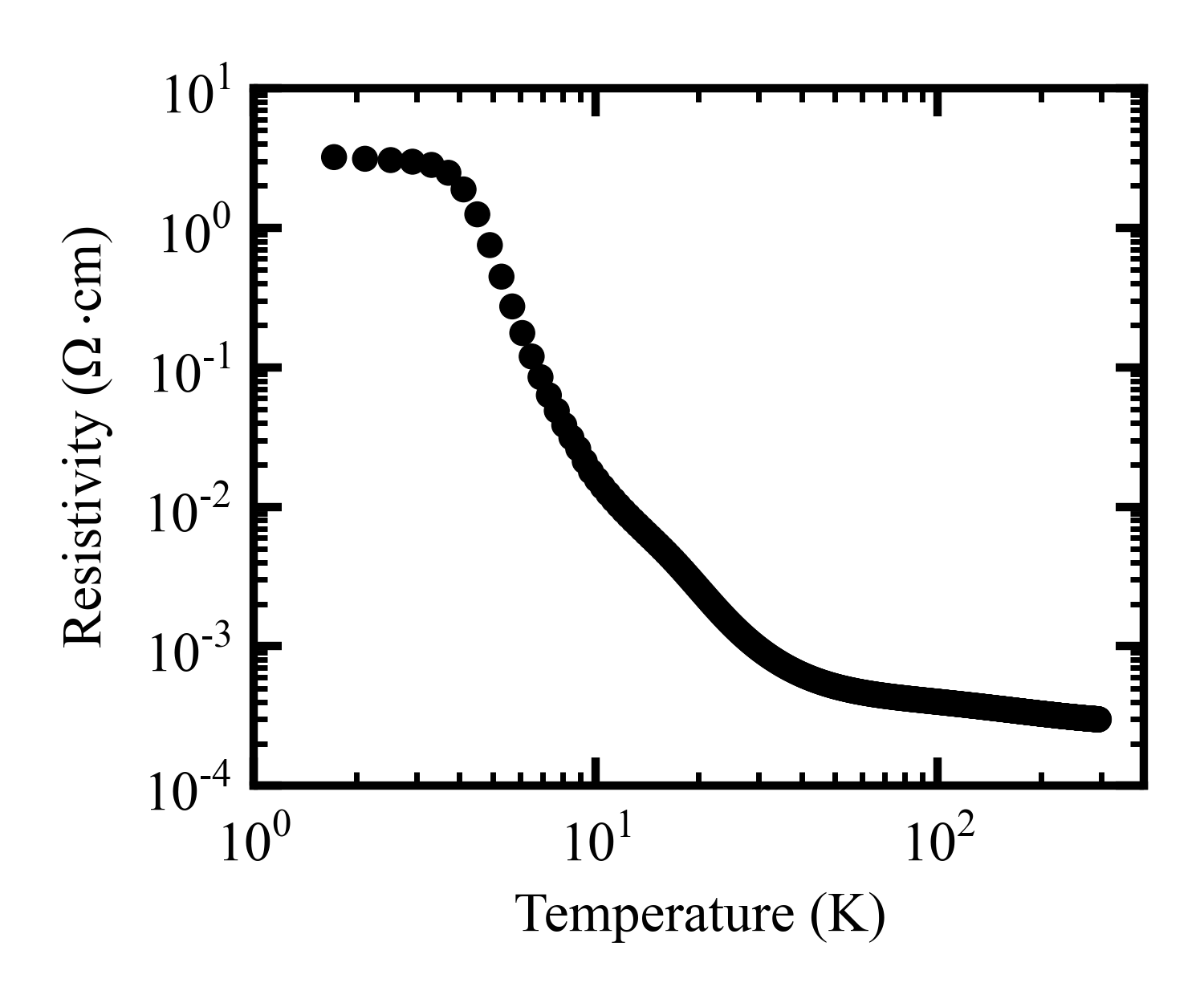}%
\caption{Resistivity vs. temperature in an Al-flux-grown \smb{} sample. Note that below 4 K, surface conduction dominates, so the plateau should not be interpreted as bulk resistivity. \label{fig1}}
\end{figure}

One notable feature of the bulk reported in the literature is the positive sign of the Hall coefficient from 65 K to room temperature. \cite{allen, molnar, nickerson} The dominant carrier in \smb{} is known to be electrons based on ARPES measurements, \cite{denlingerlinked} so the Hall coefficient would be expected to be negative at all temperatures. However, at high temperatures, the Hall coefficient is measured to be positive and appears to suggest that conduction is dominated by holes. This feature has been attributed to to skew scattering arising from the strong correlations between $f$ and $d$ electrons in a magnetic field at temperatures above where the hybridization gap opens. \cite{coleman85} In regimes where skew scattering is very strong, it can become dominant over conventional scattering mechanisms like impurity or phonon scattering, sometimes causing an inaccurate measurement of the Hall coefficient. In certain cases, including \smb, the Hall effect can no longer reliably be used to indicate the sign of the carriers. In this work we investigate the effects of the strong $f$-$d$ correlations and skew scattering on transport, extending the temperature range up to 400 K to further study the bulk properties of \smb.

\begin{figure}
\includegraphics[scale=1, trim = 0mm 0mm 0mm 0mm, clip]{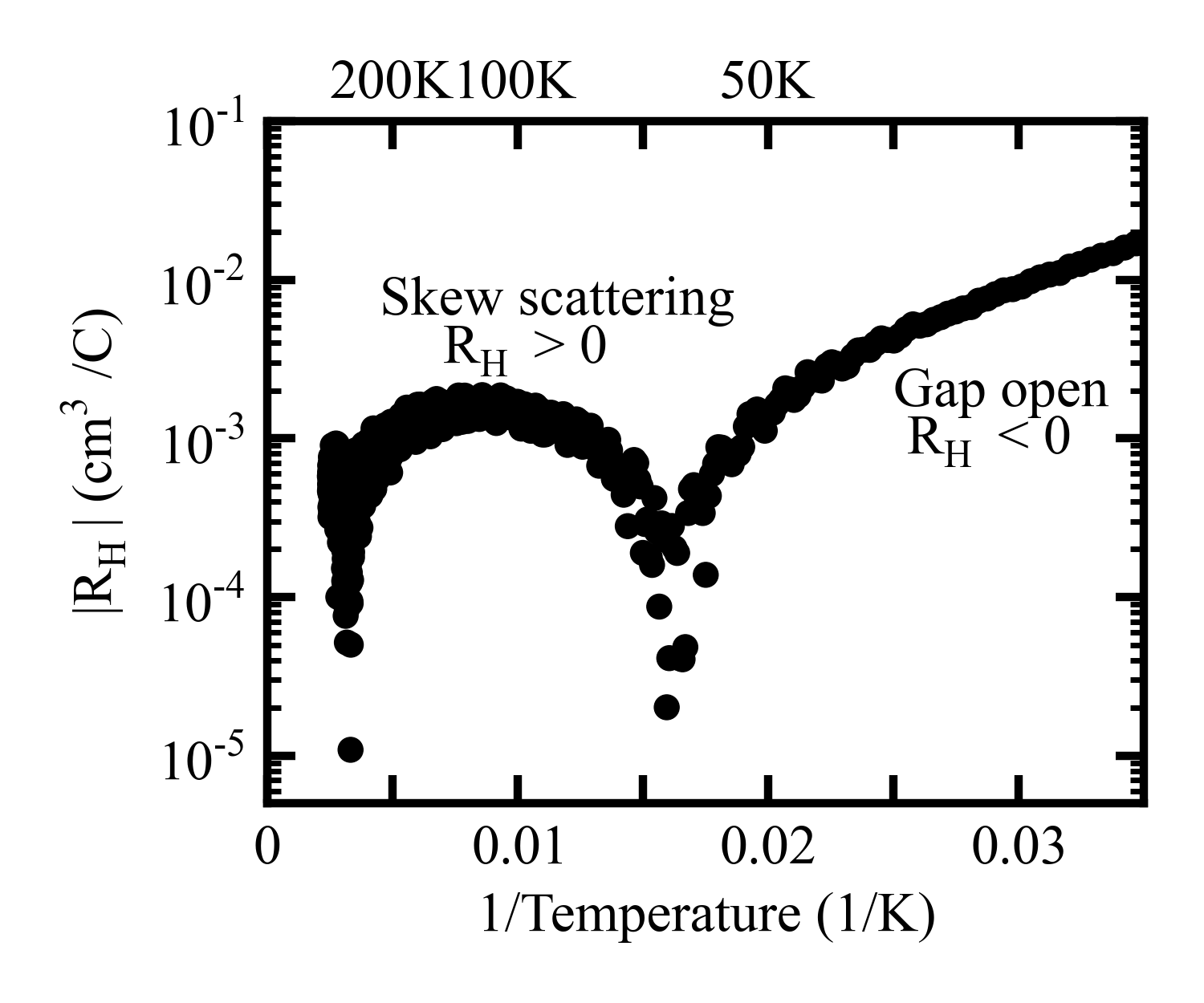}%
\caption{Hall coefficient vs. temperature in an Al-flux-grown \smb{} sample. \label{fig2}}
\end{figure}

The sample tested in this work was grown by the aluminum flux method, and before measuring, it was thinned to approximately 65 $\mathrm{\mu}$m to ensure that no Al-flux inclusions remained. Transport measurements were performed on a standard Hall bar geometry which was defined by depositing Ti/Au pads onto the sample. The measurements were performed in a 14T Quantum Design Physical Property Measurement System (PPMS).

The resistivity-temperature curve for our sample is shown in Fig. \ref{fig1}. This sample demonstrates the well-known resistance rise and plateau of \smb, indicating that it is a high quality sample. Fig. \ref{fig2} shows the Hall coefficient results at high temperatures. Our data reproduces the reported sign change feature at about 65 K; below this temperature the Hall sign is negative, and above this temperature the Hall sign is positive due to strong skew scattering. In extending our measurements above room temperature, we also observed a second sign change, with the Hall coefficient again measured to be negative above about 305 K. 

\begin{figure}
\includegraphics[scale=1, trim = 0mm 0mm 0mm 0mm, clip]{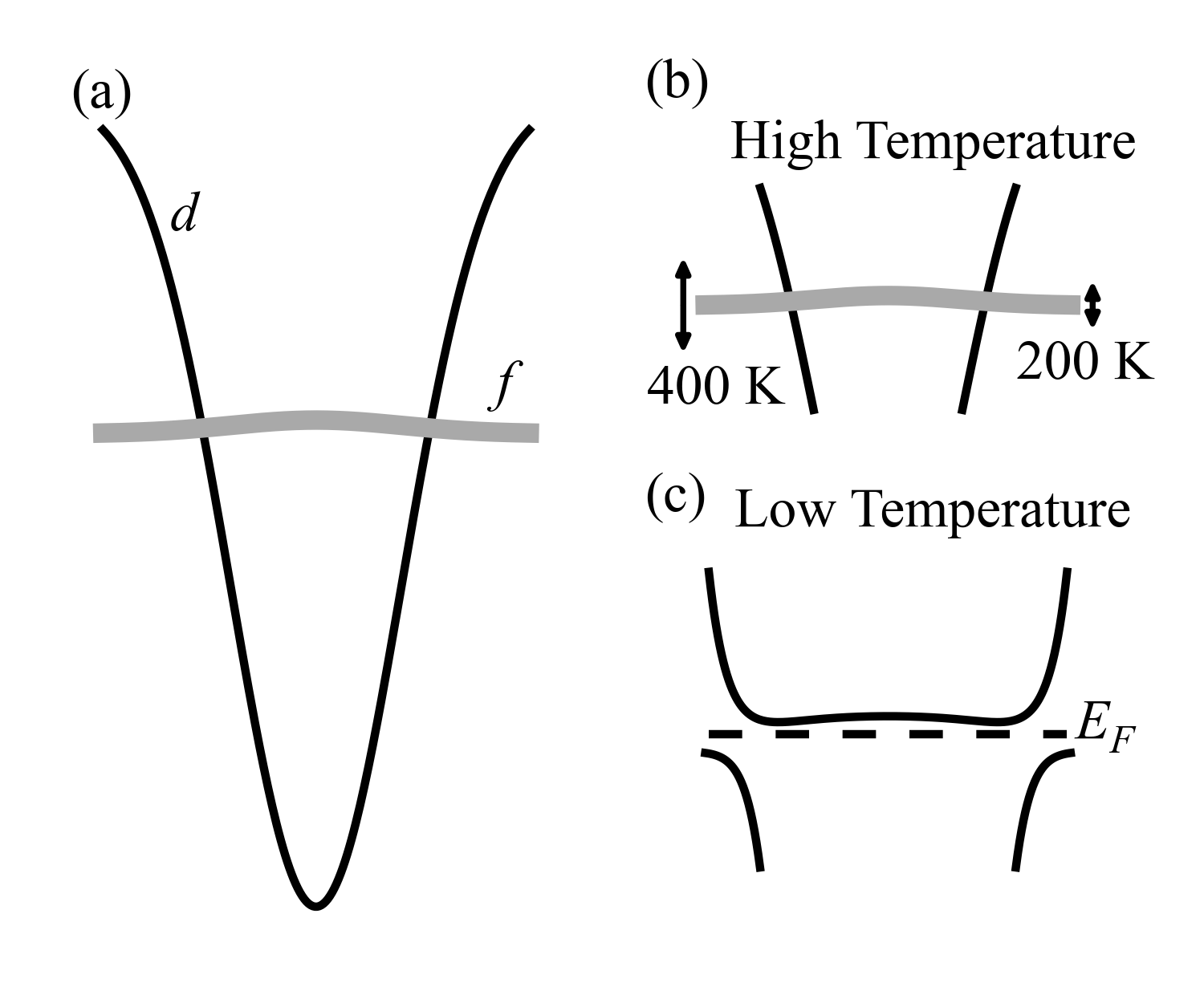}%
\caption{(a) Band diagram schematic of \smb{} at room temperature. (b) Zoom of (a) with arrows indicating $\kt$ at 200 K and at 400 K. (c) Band diagram of \smb{} at low temperatures where the gap is open.  The three diagrams are not drawn to scale, and only the $f$ band that participates in hybridization is shown. \label{fig3}}
\end{figure}

Skew scattering and its relationship to the two measured sign changes in the Hall coefficient can be understood using energy considerations for both the $f$ and $d$ electrons. To illustrate this, a schematic of the unhybridized band structure of \smb{} is shown in Fig. \ref{fig3}(a). The $f$ level that participates in hybridization has a weak $k$ dispersion as well as an intrinsic broadening, which are shown in the thick gray line in the figure. The intrinsic broadening is difficult to isolate due to the presence of thermal effects, \cite{denlingerlinked} and the $k$ dispersion is also expected to decrease with increasing temperature. Incorporating the effects of both the dispersion and the broadening, we define the overall width of the $f$ level to be $\Delta E_f$. To understand skew scattering, this overall width can be compared to the thermal energy $\kt$ of the system, because only the $d$ electrons within $\kt$ of the Fermi energy participate in transport. Since the Fermi energy is resonant with the $f$ levels, when $\Delta E_f$ and $\kt$ are comparable, the $d$ electrons that participate in transport are also resonant, leading to strong skew scattering.

These energy considerations and the effects of resonance are dependent on temperature. At the highest temperatures, $\kt \gg \Delta E_f$, meaning that few $d$ electrons are in resonance with the $f$ level. Here, skew scattering is negligible, and we observe the expected electron-dominated Hall behavior characterized by conventional scattering mechanisms like impurity or phonon scattering. This regime is illustrated in Fig. \ref{fig3}(b) with the arrow on the left, which indicates $\kt$ at 400 K. As the temperature decreases, more $d$ electrons become resonant with the $f$ level, as illustrated with the arrow on the right of Fig. \ref{fig3}(b). When resonance increases, the skew scattering effects also increase, eventually leading to the sign change to positive Hall coefficient in \smb. Skew scattering is at its strongest when $\kt \sim \Delta E_f$.  As the temperature is lowered further, the gap begins to open at the Fermi level. \cite{zhang} When the gap is open, as shown in Fig. \ref{fig3}(c), there are not enough electrons left at the Fermi level to participate in resonance. Here, the skew scattering mechanism is destroyed and the Hall coefficient is again measured with the correct sign down to lowest temperatures.  

\begin{figure}
\includegraphics[scale=1, trim = 0mm 0mm 0mm 0mm, clip]{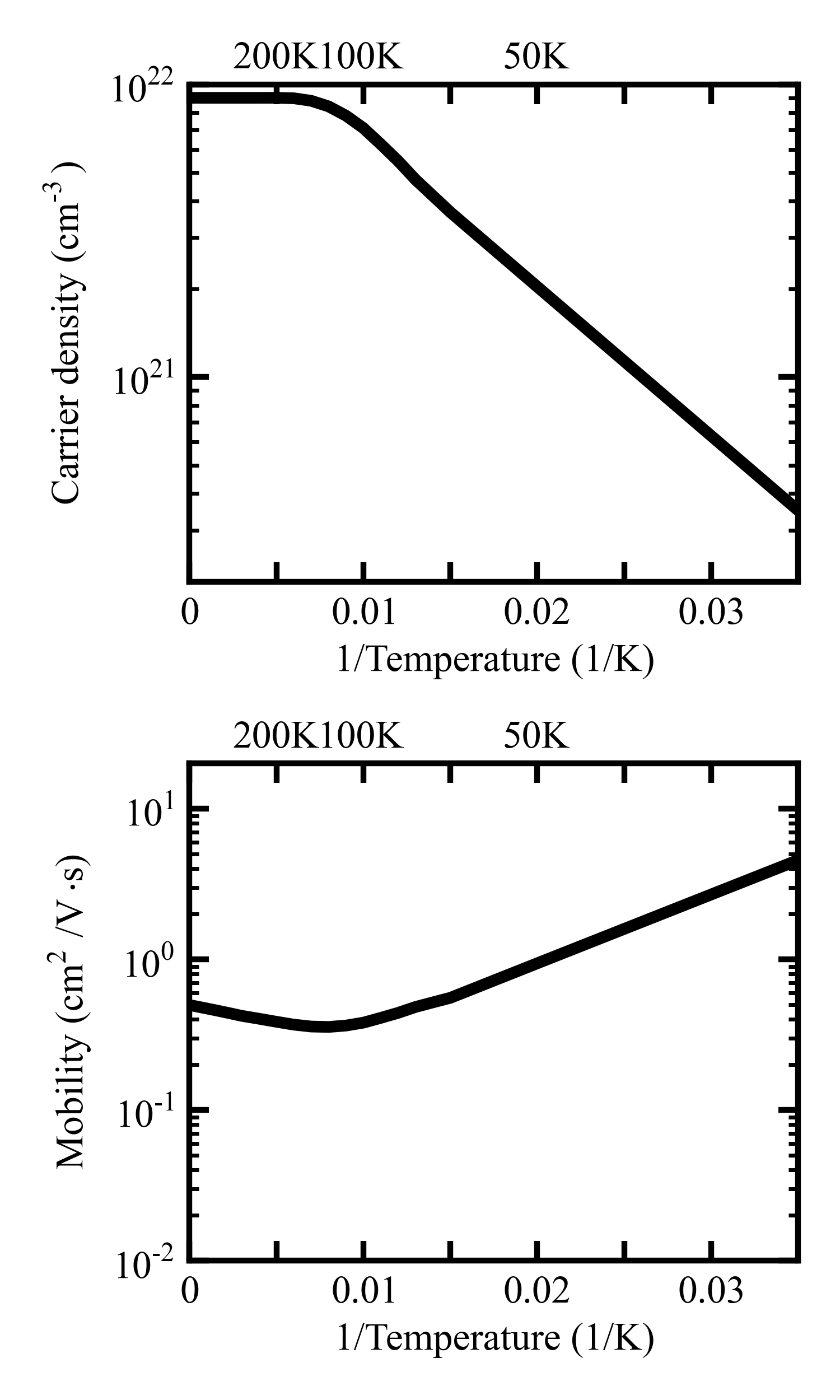}%
\caption{(a) Estimated carrier density after removing skew scattering effects. (b) Mobility calculated by combining the curve in (a) with the curve in Fig. \ref{fig1}. \label{fig4}}
\end{figure}

Since the Hall coefficient is often used to extract carrier density and mobility, the effects of skew scattering can present problems in transport analysis. The carrier density extracted from the \smb{} Hall curve ($n = 1/(eR_H)$) is inaccurate in both magnitude and sign between about 50 and 320 K, which then leads to an inaccurate determination of mobility ($\mu = 1/(ne\rho)$)  in this range as well. To rectify this, we combined transport and ARPES data to estimate the Hall coefficient over the temperature range shown in Fig. \ref{fig2}. In regions where the transport behavior is activated (below about 50 K), we can fit the Hall curve ($\rho = \rho_0 e^{\kt}$). From the Fermi pockets measured in ARPES, \cite{denlingerlinked, rakoski} we calculate that the maximum carrier density available in the Brillouin zone is $9.0 \times 10^{21}$ cm$^{-3}$. This value corresponds to the Hall coefficient at infinite temperature. These two segments can then be connected by fitting a smooth curve, yielding an estimate of the true carrier density without skew scattering, which is shown in Fig. \ref{fig4}(a). 

The carrier density can then be used to extract the mobility, as shown in Fig. \ref{fig4}(b). The features at high temperature in the mobility curve are quite different than those that would be present using the standard analysis methods. Most notably, a minimum in mobility of 3.7$\times$10$^{-1}$ cm$^2$/V$\cdot$s appears at about 100 K. Since the hybridization gap begins to open just below 100 K, \cite{zhang} it is expected that the effects of skew scattering are strongest near this minimum. By combining the minimum mobility with the effective mass and Fermi velocity extracted from ARPES data, \cite{denlingerlinked,denlingerpe} we obtain a scattering time of 0.18 fs and a mean free path of 1.1 \AA. These results also satisfy $k_F l < 1$, confirming that the scattering near the minimum in the mobility is very strong. At room temperature, for comparison, the mobility is 4.2$\times$10$^{-1}$ cm$^2$/V$\cdot$s, which corresponds to a scattering time of 0.21 fs and a mean free path of 1.3 \AA. The increase in mobility is due to the higher thermal energy of scatterers at room temperature compared to 100 K, meaning that fewer $d$ electrons are resonant with $E_f$ at the higher temperature.

In this work, we have shown that Hall coefficient data demonstrates an incorrect positive sign from 65 K to 305 K. We attribute this feature to skew scattering, which is the dominant scattering mechanism in this temperature range. Skew scattering arises when a large fraction of the mobile $d$ electrons are in resonance with the $f$ electrons at the Fermi energy, so the effect is strongest at these intermediate temperatures. At temperatures above 305 K, the large thermal energy of the $d$ electrons leads to little resonance with the $f$ electrons, and the Hall coefficient sign is measured correctly. At around 100 K, the hybridization gap begins to open, destroying the skew scattering effect as the temperature continues to decrease. Below 65 K, skew scattering is no longer dominant and the Hall coefficient sign is again correct.

The effects of skew scattering lead to an incorrect determination of the carrier density and mobility, so we used parameters extracted from ARPES data to ``patch'' together the region where skew scattering dominates. We obtained an estimate for the carrier density without skew scattering as well as the corresponding mobility. The mobility results show a minimum around about 100 K, which is where skew scattering is the strongest. This method allows the correct transport carrier density and mobility to be obtained for \smb. Although this work is a study of the \smb{} bulk, much of the modern work focuses on the interesting surface characteristics of \smb. Both surface and bulk characteristics will likely continue to be studied heavily in the context of the relationship between the strong electronic correlations and topological effects in this unique material.

\begin{acknowledgments}
The authors would like to thank K. Sun for helpful discussions in preparing this work. Device fabrication was performed in part at the University of Michigan Lurie Nanofabrication Facility. Funding for this work was provided by NSF grant \#DGE-1256260. 
\end{acknowledgments}

\bibliography{allsmb6papers}

\end{document}